# ChatGPT and Excel – trust, but verify

Patrick O'Beirne

Systems Modelling Ltd

Ireland

pob@sysmod.com

**ABSTRACT**

This paper adopts a critical approach to ChatGPT, showing how its huge reach makes it a useful tool for people with simple requirements but a bad, even misleading guide to those with more complex problems which are more rarely present in the training data and even more rarely have straightforward solutions.

It works through four exercises in creating lookup formulas using chatbots, showing the need to test the offered solutions. They are a simple lookup, a lookup to the left, a match of two values at the same time, and intentionally confusing the models by using common language with technical meaning in Excel.

It concludes with a practical guide for how to add an Excelscript button, with system and user prompts, to the ChatGPT API into the Excel desktop environment, supported by a blog post giving the technical details for those interested.

## 1       OVERVIEW

We review the subject of Spreadsheet Risk and introduce the notion of Large Language Models (LLM's). They are attracting increasing attention in multiple domains and discuss the current state of the art and the use of a particular LLM in the field of accountancy, where spreadsheets have a dominant application. We discuss Excel formulas and VBA through a series of video links and related successes and failures together with Security concerns. We show how to add a ChatGPT query box to Excel. We conclude with Ten lessons learned and a summary. In the Appendix we document four exercises creating lookup formulas using chatbots.

### 1.1      Trust issues in spreadsheets

Spreadsheets are everywhere, used from quick scratchpads to planning decisions, operational processes, valuations, financial reporting, and analysis. Excel is the "Swiss Army knife" for data analysis and calculation. But however qualified users are in their own domain, few have received training in Excel to the level they should have to use it efficiently and effectively in their work. Because they don't know what they don't know, they rate themselves overconfidently as "proficient in Microsoft Excel". This gives rise to what has been called the "Sorcerer's apprentice" syndrome, named from a story where an apprentice magician knew enough to cast a spell, but not enough to stop it when it ran out of control.

The EuSpRIG horror stories page [EuSpRIG, 2020] has many stories of the expensive and embarrassing results of out-of-control spreadsheets, such as the JP Morgan $6 Billion trading loss, and the Reinhart and Rogoff 'austerity spreadsheet' error. Data quality issues

are frequently reported, for example the articles in scientific journals about the widespread occurrence of gene name errors caused by Excel misinterpreting names such as SEPT2 as a date. As a result, researchers have now renamed genes to avoid this problem.

A large body of research has been published by the European Spreadsheet Risk Interest Group (EuSpRIG, 2020) on the risks in using spreadsheets, stemming from user under-training but over-confidence. Trust issues in automated solutions have been researched by Muir & Moray [1994] and Croll [2017]. Many online forums exist to support users of Excel and they represent a very large resource of knowledge. Millions of users trawl these every day either spending hours Googling and sifting the useful information from the spam, or posting their current Excel challenges and receiving answers, sometimes choices of answers, from Excel experts such as Microsoft Valued Professionals (MVPs). Now they have new machine-learning (branded "AI") tools with vastly more patience and capacity than they have to scrape the millions of existing questions and answers. These are the large language models (LLMs) such as ChatGPT-4 which are behind ChatGPT, Microsoft Bing and Copilot [Microsoft, 2023], and Google Bard.

### 1.2 Large Language Models

A Large Language Model (LLM) is an algorithm has been trained on a large amount of text-based data, typically scraped from the open internet, and so includes billions of web pages and - depending on the LLM - other sources such as scientific research, books or social media posts. Recent examples of LLMs include OpenAI's ChatGPT, Microsoft Bing, Google's Bard and Meta's LLaMa (for scientific papers).

They are very good at reproducing the syntax of sentences in natural languages but have no concept of semantics, the meaning of the words. But that does not stop people attributing intelligence and understanding to them. Some people are now calling themselves "Prompt Engineers". Prof. Ethan Mollick of Wharton University has a blog [Mollick, 2023] on how to optimize the use of ChatGPT and other LLMs.

In the domain of spreadsheets, or code more generally, testing is the most important step, because users of ChatGPT necessarily do not have the skills to create the formula themselves, or even to understand how the offered formulas work, so all they can do is test.

### 1.3 Current state of the art

For common problems with published solutions, the suggestions are correct most of the time. That reflects the weight of such Q&As in the sources. When there is less information to scrape, ChatGPT makes stuff up based on whatever it can find on similar topics. When asked to create a list of the ten most cited papers on a research topic, the first two were close to real papers, the rest were completely invented. I tested both ChatGPT and Bing by giving each the top five papers and asking them to check whether they were genuine by providing links to online sources. ChatGPT invented non-existent links, or links to completely different papers, and created more different but non-existent links when the response was regenerated. Bing gave a link for one, a correct paper, and declined to provide links to the rest, saying that they were either not genuine or behind a paywall.

Bing found two different papers and said that it could not find more. Bing is less creative and more factual. It gives links to search results to learn more. Microsoft have announced Copilot, a formula assistant in Excel [Microsoft, 2023].

In an article on the website of the Institute of Chartered Accountants of England & Wales (ICAEW) "ChatGPT and accountancy: know its limits" [ICAEW, 2023], Stuart Cobbe put it through the ICAEW assurance assessment paper. The version in place at the time, ChatGPT-3.5, failed with a score of 42%. When ChatGPT-4 was released, Cobbe tried the assessment again. This time, it passed with 78% – a significant improvement on last time, with better understanding and reduced rambling. You should consider it, says Cobbe, "like an ill-informed but confident junior."

### 1.4  ChatGPT successes

Mynda Treacy [Treacy, 2023] pasted in a question on Power Pivot from her forum, and it gave the correct answer using the column names in the question.

I tried a question from the Excel-L list: Tell me why I get an error "Syntax error in INSERT INTO statement." in the following SQL statement:

INSERT INTO stagingHatchCommList ( Type, Direction, To, ContactFrom,…

Answer: "may be caused by the use of reserved keywords as column names"

"Use INSERT INTO stagingHatchCommList ( [Type], [Direction], [To], …"

Paul Kelly [Kelly, 2023] asked it to create a VBA Class module, a a wrapper class for a Collection that implements a Reverse method. The answer impressed him.

### 1.5  ChatGPT failures

I asked it for a formula to convert IP addresses with four 3-digit parts that have been misinterpreted as numbers, eg 121451781090, into the correct form, eg 012.145.178.090

ChatGPT offered several solutions, none of which worked.

One solution is SUBSTITUTE(TEXT(A7,"000|000|000|000"),"|","."); another uses a custom number format, =TEXT(A7,"000\.000\.000\.000").

### 1.6  Security concerns

The UK National Cyber Security Centre (NCSC) [UK NCC, 2023] reports that "…concerns that the OpenAI tools could turn any would-be petty thief into a technically savvy malicious coder in moments were, and still are, well-founded. ChatGPT-enabled cyberattacks started popping up just after its user-friendly interface premiered in November 2022. …  For more complex tasks, it's currently easier for an expert to create the malware from scratch, rather than having to spend time correcting what the LLM has produced. However, an expert capable of creating highly capable malware is likely to be able to coax an LLM into writing capable malware."

They advise:

Do not include sensitive information in queries to public LLMs

Do not submit queries to public LLMs that would lead to issues were they made public

Add a ChatGPT-3.5 query box to Excel

The Author's blog [O'Beirne, 2023] shows a way of building an interface to the OpenAI ChatGPT-3.5 model.

It uses Excelscript (a variety of Typescript) so sits in your Excel desktop application, not in a workbook, so it does not need VBA. (VBA macros can be found as well, of course [Mullaned 2002, 2023])

You can download the ChatGPT in Excel demo workbook from [O'Beirne, 2023]

## 2    SUMMARY

### 2.1    Lessons Learned

The ten lessons learned from the experiments in the Appendix are:

> 1: Test the formula suggested, even if it looks right to you – or in other words, you don't see any problem at first.
>
> 2: If it does not work, try again. Give it more context, say what does not work.
>
> 3: Start over again with a fresh session and ask the same question to see if you get the same answer.
>
> 4: If you know of a newer function that you think might work, suggest it in your prompt.
>
> 5: When you copy down a formula, ensure that the references are correct in both relative and absolute forms.
>
> 6: Do more than one test and verify each result.
>
> 7: It's best to really understand how the formula works so that you can decide whether it is a good match for your needs.
>
> 8: When you get a formula, ask what could go wrong with it and check each potential issue.
>
> 9: When you get a formula you can't understand, ask a human expert to explain it.
>
> 10: Remember, these are language models, not logical or mathematical models.

### 2.2    In conclusion – trust, but verify

Use LLMs for questions where there should be a known solution. Verify any statements against trusted sources. Bing provides links to known sources. Test any code (Excel formulas, VBA, or other programming code) with realistic data and ensure that the results are correct before putting it to use. Repeat and refine the suggestions to see if you get a better answer.

# APPENDIX

Experiments in ChatGPT and Microsoft Bing with Excel formula questions

In the dialogues following, I have edited the questions and verbose responses to save space. I chose to try lookup exercises.

Exercise 1: a simple lookup

> I have a list of products and prices in A10:B110, and a list of the products sold and their sales quantity in J11:K50. In column L I want a calculation of the quantity times the price for each product.

| | A | B | C | J | K | L |
|---|---|---|---|---|---|---|
| 9 | | | | | | |
| 10 | productname | Price | | Product | Qty | value |
| 11 | Beef - Rib Eye Aaa | 6.98 | | Sole - Fillet | 10 | 197.9 |
| 12 | Corn Syrup | 7.78 | | Brandy - Orange, Mc Guiness | 18 | |
| 13 | Sesame Seed Black | 10.61 | | Pork Loin Cutlets | 20 | |
| 14 | Cheese Cloth | 2.7 | | Chocolate Bar - Smarties | 14 | |
| 15 | Sole - Dover, Whole, Fresh | 15.38 | | Shiro Miso | 6 | |
| 16 | Sauce - Caesar Dressing | 1.4 | | Island Oasis - Sweet And Sour Mix | 10 | |
| 17 | Cheese - Augre Des Champs | 1.4 | | Lamb - Shoulder | 12 | |
| 18 | Shrimp - Baby, Cold Water | 15.29 | | Shrimp - Baby, Cold Water | 8 | |
| 19 | Sole - Fillet | 19.79 | | Bandage - Fexible 1x3 | 6 | |
| 20 | Wine - Sauvignon Blanc Oyste | 5.56 | | Creme De Banane - Marie | 18 | |

L11: =VLOOKUP(J11,$A$10:$B$110,2,FALSE)*K11

**Figure 1: simple lookup example with one formula completed**

ChatGPT-3 was correct: =VLOOKUP(J11,$A$10:$B$110,2,FALSE)*K11

The VLOOKUP function examples returned all specified the fourth parameter [range_lookup] correctly as FALSE, which does not require the lookup list to be sorted in ascending order and therefore uses a linear search from the beginning of the list. They did not make the common user mistake of omitting it; the default value for the omitted parameter is TRUE, which requires the lookup list to be sorted so that a fast binary split search can be used.

ChatGPT-3.5 first got the columns the wrong way around: =J11*VLOOKUP(K11,$A$10:$B$110,2,FALSE). When told that, it corrected it. A naive user might just type it in without noticing that and would get a #VALUE! result. Even if they did notice it, they may still decide to do it as a test.

**Lesson 1: Test the formula suggested, even if it looks right to you – or in other words, you don't see any problem at first.**

Bing made the same mistake: =J11*VLOOKUP(K11,$A$10:$B$110,2,FALSE). When told "The product name is in column J and the quantity in column K", it responded "Thank you for letting me know. …. =K11*VLOOKUP(J11,$A$10:$B$110,2,FALSE)".

Exercise 2: reverse lookup, ie lookup to the left

I then tried a swapping the columns for product name and price, asking for separate formulas for lookup and multiplication.

*I have a list of products in C11:C110, and prices in B11:B110, and a list of the products sold in J11:J50 and their sales quantity in K11:K50. In column L I want the price of each product sold and in column M a calculation of the quantity times the price for each product.*

| B | C | DEFGHI | J | K | L | M |
|---|---|---|---|---|---|---|
| 10 Price | productname | | Product | Qty | price | value |
| 11  6.98 | Beef - Rib Eye Aaa | | Cheese Cloth | 2 | 2.7 | 5.4 |
| 12  7.78 | Corn Syrup | | Brandy - Orange, Mc Guiness | 18 | 11.19 | |
| 13  10.61 | Sesame Seed Black | | Pork Loin Cutlets | 20 | 18.29 | |
| 14  2.7 | Cheese Cloth | | Chocolate Bar - Smarties | 14 | 17.32 | |
| 15  15.38 | Sole - Dover, Whole, Fresh | | Shiro Miso | 6 | 0.6 | |

L11 formula: =INDEX(B10:B110,MATCH(J11,C10:C110,0))

**Figure 2: more complex lookup with the lookup column after the returned column**

They all got the simple multiplication correct in M12, so I will focus on the lookup formula in L12.

ChatGPT-3.5 was incorrect at first: =VLOOKUP(J11,$C$10:$B$110,2,FALSE)

This formula is not correct syntax because C is after B. If you entered that in Excel, it would autocorrect it to =VLOOKUP(J11,$B$10:$C$110,2,FALSE). When I asked " What is wrong with this Excel formula: =VLOOKUP(J11,$C$10:$B$110,2,FALSE)" it correctly replied " The range $C$10:$B$110 is in the wrong order. It should be $B$10:$C$110."  Later, I tested more variations on this, see problem #4 below.

Worse than that, the formula does not work because it is looking up the product name in column B which is the product price. Applying lesson 1 and testing that, I get a #N/A value. I therefore revised the prompt by adding "*Note that the product prices are to the left of the product names so I cannot use VLOOKUP.*"  It then offered: =INDEX($B$11:$B$110,MATCH(J11,$C$11:$C$110,0)) which is correct. Sometimes simply repeating the prompt will get it to offer a different answer, but it's better to be more specific.

**Lesson 2: If it does not work, try again. Give it more context, say what does not work.**

ChatGPT-3.5 gave the same incorrect formula: =VLOOKUP(J11,$C$10:$B$110,2,FALSE). When I replied that that have a #N/A, it responded "I apologize for the mistake. The formula for column L should use the price column as the first argument and the product column as the second argument in the VLOOKUP function. Here's the corrected formula for column L: VLOOKUP(J11,$B$10:$C$110,2,FALSE) This formula looks up the price of the product sold in column J, in the B10:C110 table, and returns the value in the second column (price)." On being told again that this produced a #NA, it offered: =INDEX($B$10:$B$110,MATCH(J11,$C$10:$C$110,0)).  Repeatedly saying only "that is wrong" is a slow way of being offered an alternative, it is better to provide a reason if you know what it is; which of course naïve "copy & paste" Excel users may not.

Bing made the same VLOOKUP mistake, which I am not re-quoting here, for economy. On the third attempt, it offered for L11: "=INDEX(B10:B110,MATCH(J11,C10:C110,0)).

This will look up the value in J11 (the product sold) in column C and return the corresponding value from column B (the price). You can then drag this formula down to fill column L."

Unfortunately, when I copy that down, the relative references adjust so in M12 the formula reads =INDEX(B11:B111,MATCH(J12,C11:C111,0)), and so on down, until eventually the answers are #N/A because the references are far out. Would a naïve user think to check the copied cells, and wonder why the #N/A values appear? To know what the problem was, they would have to understand relative and absolute addressing.

In this case, I repeated the prompt in a new Bing topic and it gave the correct formula: "In cell L11, you can enter the formula =INDEX($B$11:$B$110,MATCH(J11,$C$11:$C$110,0))" It also cited four internet sources, which ChatGPT cannot do. For example, "How to use INDEX and MATCH | Exceljet. https://exceljet.net/articles/index-and-match "

Because these are probabilistic language models, they will always suggest the most frequently occurring solutions found in their sources, in this case VLOOKUP, even when they are not appropriate to the specific question. When told not to, they will then fall back to the next most common solution, INDEX/MATCH combination.

I tried telling it to use LOOKUP rather than VLOOKUP. I knew that this was an inappropriate function because LOOKUP requires the lookup range to be sorted in ascending order, but wanted to see what it would do. I expected to get =LOOKUP(J11, $C$11:$C$110, $B$11:$B$110). ChatGPT-3.5 offered the following instructions, to which I have added my comments

| ChatGPT-3.5 | My analysis |
|---|---|
| Prompt: In an Excel sheet I have a list of all product prices in B11:B110, all product names in C11:C110, and another list of the names of the products sold in J11:J50. Neither list is sorted in any order. In column L I want an Excel formula to return the price of each product sold. Note that the product prices are to the left of the product names so I cannot use VLOOKUP, I want to use LOOKUP. | |
| Response: To use LOOKUP in this scenario, you need to rearrange the data so that the product names are in the left column and the prices are in the right column. Here are the steps: | We don't need to, we could use C as the lookup vector (ie the list to look up) and B as the result vector (ie the list to take the result from) |
| | |
| 1. Select cells A11:B110. | But the data is in B11:C110 |
| 2. Cut the data (Ctrl+X or right-click and select Cut). | |
| 3. Right-click cell A11 and select Transpose. | There is no Transpose option available after a cut, only after a copy. Assume for a test that we choose Copy and a different output cell temporarily, then |

| | choose Paste Special, Transpose. That transposes the two columns to two rows. |
|---|---|
| 4. Right-click cell A11 again and select Paste Special. | |
| 5. In the Paste Special dialog box, select Transpose and click OK. | That would just repeat the previous paste and would have no effect. If we copied again, chose another destination, and copy & paste transposed, that merely gives us back our original columns, they are not swapped. |
| 6. Delete the original data in B11:C110. | |
| 7. Rename column A to "Product Name" and column B to "Price". | |
| Now that the data is rearranged, you can use LOOKUP to find the price of each product sold. Here's the formula for cell L11: | |
| =LOOKUP(J11,$A$11:$A$50,$B$11:$B$50) | The lookup and return areas should be 100, not 50. |
| | The LOOKUP function requires the lookup list to be sorted which it is not. ChatGPT doesn't know that, of course as I didn't say so. |
| This formula looks up the value in cell J11 (the product name) in the range A11:A50 (the product names) and returns the corresponding value from the range B11:B50 (the prices). You can copy this formula down to cells L12:L50 to get the prices for all the products sold. | |

When I tried again, it gave a different method of constructing rearranged columns using INDEX/MATCH, then the same LOOKUP function. INDEX/MATCH would work, but it would be an unnecessary complication.

I also tried explicitly saying that the lists were not sorted and asked if I could use LOOKUP rather than requiring it, as follows:

"In an Excel sheet I have a list of all product prices in B11:B110, all product names in C11:C110, and another list of the names of the products sold in J11:J50. Neither list is sorted in any order. In column L I want an Excel formula to return the price of each product sold. Note that the product prices are to the left of the product names so I cannot use VLOOKUP. Can I use LOOKUP?"

The response from ChatGPT3.5 was that I could, although we know that this cannot work with unsorted lists. At least it did not tell me to rearrange the data:

"Yes, you can use the LOOKUP function to return the price of each product sold. Here's the formula you can use in cell L11: =LOOKUP(J11,C$11:C$110,B$11:B$110)

This formula will search for the product name in cell J11 in the range C11:C110, and return the corresponding price from the range B11:B110. You can then copy this formula down to cells L12:L50 to get the prices for all the products sold."

Bing responded: " No, you cannot use the LOOKUP function in this case because the lookup value must be in the first column of the array. Instead, you can use the INDEX and MATCH functions together to achieve the desired result. In cell L11, you can enter the formula `=INDEX(B11:B110,MATCH(J11,C11:C110,0))`". This is a correct solution for J11, although the reasoning about LOOKUP was wrong. It would not work copied down unless the C and B references were absolute.

I then asked about XLOOKUP. This is a new function in Office 365 and so will appear in fewer web searches and therefore have a low probability of being suggested by a LLM. This worked but no response used absolute references in the lookup table.

ChatGPT-3.5 responded " "Yes, you can use XLOOKUP to return the price of each product sold. Here's the formula you can use in cell L11: =XLOOKUP(J11,C11:C110,B11:B110)". This is correct for J11 (but not suitable for copying down because not absolute) even though the optional match_mode parameter was not specified, because the XLOOKUP default is an exact match.

Bing gave the same relative-only formula and helpfully added "Please note that XLOOKUP is not available in Excel 2016 and Excel 2019. If you have an older version of Excel, you may need to use a different function such as INDEX and MATCH." It gave links to support.microsoft.com and other sites.

**Lesson 3: Start over again with a fresh session and ask the same question to see if you get the same answer.**

**Lesson 4: If you know of a newer function that you think might work, suggest it in your prompt.**

**Lesson 5: When you copy down a formula, ensure that the references are correct in both relative and absolute forms.**

Exercise 3: Lookup to match two values and return the last match

This is moderately difficult because it requires matching multiple rows in a table and returning the last result that satisfies the criteria. It requires a deeper knowledge of Excel to understand, test, and debug such formulas.

Here is the lookup table, a list of several price changes on different dates for each of three products:

|   | A | B | C | D |
|---|---|---|---|---|
| 1 | ID | Productname | Date | Price |
| 2 | 171 | Apple - Delicious, Red | 11-Mar-20 | 17.36 |
| 3 | 178 | Appetizer - Lobster Phyllo Roll | 18-Mar-20 | 15.27 |
| 4 | 199 | Apple - Granny Smith | 08-Apr-20 | 2.70 |
| 5 | 171 | Apple - Delicious, Red | 01-Jan-21 | 6.58 |
| 6 | 178 | Appetizer - Lobster Phyllo Roll | 01-Jan-21 | 16.24 |
| 7 | 199 | Apple - Granny Smith | 01-Jan-21 | 14.98 |
| 8 | 171 | Apple - Delicious, Red | 12-Mar-23 | 2.58 |
| 9 | 178 | Appetizer - Lobster Phyllo Roll | 19-Mar-23 | 15.39 |
| 10 | 199 | Apple - Granny Smith | 09-Apr-23 | 16.67 |

**Figure 3 – a price change history table**

My first prompt did not specifically say that there would be more than one price for a given product:

"I have in Excel a table with product price history from A2 to D10. Column B is the list of products, column C the dates that the product price was changed, and column D is the price set on that date. I want to find the price in column D for a given product name in G2 and a given sales date in H2."

So the first formula that both ChatGPT and Bing offered would only work if there was only one price per product, as it returns the first exact match:

=VLOOKUP(G2, B2:D10, 3, FALSE)

I then specified it more carefully by adding "The sales date can be any date in between two price changes." It still offered VLOOKUP but even more incorrectly:

"To find the price of the product sold on a given date, you can use the following formula: =VLOOKUP(G2,A2:D10,4,TRUE) This formula uses the VLOOKUP function to search for the product name in column A and return the corresponding price from column D. The "TRUE" argument in the formula tells Excel to find an approximate match for the sales date in column C, which means it will return the price for the most recent date before the sales date."

I tried again, being even more explicit:

"I have in Excel a table with product price history from A2 to D10. Column B is the list of products, column C the dates that the product price was changed, and column D is the price set on that date. Given a product name in G2 and its sales date in H2, I want to find the price in the product history for the given product name in G2 and the latest date not on or after the given date in H2."

Now both ChatGPT (and ExcelFormulaBot.com which uses ChatGPT) and Bing offered an INDEX/MATCH formula.

=INDEX(D2:D10,MATCH(1,(B2:B10=G2)*(C2:C10<=H2),0))

However, that is not correct and their explanation of how it worked was not correct. I have put that in bold font. Because it finds the **first** 1, it actually returns the first date before the given date, not the last. I have left the rest of the explanation there for readers who wish to understand how to lookup an array of calculated Boolean values from multiple conditions combined by AND.

"This is an array formula, so you need to press Ctrl + Shift + Enter instead of just Enter after typing it in. Here's how the formula works:

- MATCH(1,(B2:B10=G2)*(C2:C10<=H2),0) returns the row number of **the first match for the given product name in G2 and the latest date not on or after the given date in H2.** The formula uses two conditions: (B2:B10=G2) checks if the product name in each row matches the given product name in G2, and (C2:C10<=H2) checks if the date in each row is not on or after the given date in H2. The two conditions are multiplied together using the * operator, which returns an array of 1's and 0's. **The MATCH function then looks for the first 1 in the array and returns its position.**

- INDEX(D2:D10,MATCH(1,(B2:B10=G2)*(C2:C10<=H2),0)) returns the value in column D at the row number returned by the MATCH function. This gives you the price for the given product name in G2 and the latest date not on or after the given date in H2."

I tried yet again, adding to the prompt "There can be multiple price change entries for each product."

Now ChatGPT supplied a formula that checked for the date being the maximum date of change in the list for the given product and dates on or before the sales date:

=INDEX(D2:D10,MATCH(1,(B2:B10=G2)*(C2:C10<=MAX(IF(B2:B10=G2,C2:C10,H2+1)))),1)

This, as is often said, "looks right". And in fact, a first test gives the correct answer for the Appetizer product. But not for the Granny Smith.

```
=INDEX($D$2:$D$10,MATCH(1, ($B$2:$B$10=G2) *
($C$2:$C$10<=MAX(IF($B$2:$B$10=G2,$C$2:$C$10,$H$2+1)))),1)
```

| G | H | I | J |
|---|---|---|---|
| Product | SalesDate | Price | |
| Appetizer - Lobster Phyllo Roll | 01/01/2022 | 16.24 | Correct, lucky |
| Apple - Granny Smith | 01/01/2022 | 16.67 | Incorrect, should be 14.98 |

Figure 4 – an incorrect solution

To understand why, you need to see among all those brackets that there is no match_mode specified for the MATCH function, so it defaults to a binary split search for a 1 in an array of mixed 0s and 1s.

=MATCH(1,{0;1;0;0;1;0;0;1;0})

I leave it as an exercise for the reader to manually practice such a binary search in each case and see how it lands on essentially a random position depending on the unsorted array.

**Lesson 6: Do more than one test and verify each result.**

**Lesson 7: It's best to really understand how the formula works so that you can decide whether it is a good match for your needs.**

Those two lessons are perhaps the reason why many spreadsheets created with the assistance of LLMs will be bug-ridden because users who only know how to copy & paste such formulas do not understand how they work.

I asked "What could go wrong with this Excel formula: =INDEX(D2:D10,MATCH(1,(B2:B10=G2)*(C2:C10<=MAX(IF(B2:B10=G2,C2:C10,H2+1)))),1)"

ChatGPT said "As an AI language model, I cannot see the data in the cells, so I cannot determine if the formula is correct or not. However, here are some potential issues…" and gave five issues including "3. If the data in the ranges B2:B10 and C2:C10 are not sorted in ascending order, the formula may not return the correct result."

**Lesson 8: When you get a formula, ask what could go wrong with it and check each potential issue.**

ChatGPT-3.5 first offered an ingenious solution: interpolation. I didn't pursue it.

"To find the price at which the product would have been sold on a given sales date, you need to look up the two dates between which the sales date falls and then use linear interpolation to calculate the price. Here's a step-by-step guide to doing this: Sort the table by the dates in column C in ascending order to make sure that the dates are in chronological order. Use the MATCH function to find the row number of the last date in column C that is less than or equal to the sales date. =MATCH(H2,C:C,1) Use the INDEX function to get the date of the most recent price change before the sales date……" . I asked again and got

=INDEX(D:D,MATCH(1,(B:B=G2)*(C:C<=H2),0)+MAX((B:B=G2)*(C:C<=H2)*(C:C)))

(B:B=G2)*(C:C<=H2) gives {0;0;1;0;0;1;0;0;0;0}

MATCH(1, {0;0;1;0;0;1;0;0;0;0}, 0) returns the first exact match of 1, which is row 3 so that's the FIRST date the product price was changed on; not the last date before the sought date.

((B:B=G2)*(C:C<=H2)*(C:C) returns {#VALUE!;0;43908;0;0;44197;0;0;0;0} because there is a header label row 1. The MAX, intended to find the highest valid date, would therefore return #VALUE! and the only way a user would know why is to perform the array calculation as I did to see where the error value came from.

Even if that did work by removing the header label, adding 3+44197 would give a number 44200 as the INDEX row, which is completely wrong.

By the way, each time ChatGPT says "This is an array formula, so you need to press Ctrl+Shift+Enter instead of just Enter when entering it", that advice may not apply in modern Excel with dynamic array formulas. It doesn't do any harm, though.

**Lesson 9: When you get a formula you can't understand, ask a human expert to explain it.**

ChatGPT-3.5 gave this formula, with an explanation. It returns the price in the row containing the maximum date that satisfies both the product and date criteria.

=INDEX(D:D,MAX(IF((B:B=G2)*(C:C<=H2),ROW(C:C))),1)

If a range is chosen not starting in row 1, the index number needs to be reduced accordingly. If starting in row 2, take 1 off:
=INDEX(D2:D10,MAX(IF((B2:B10=G2)*(C2:C10<=H2),ROW(C2:C10)))-1,1)

So finally, a correct answer is produced. I retried ExcelFormulaBot saying that there were multiple dates for a product, and it offered:

=LOOKUP(2,1/((B2:B10=G2)*(C2:C10<=H2)),D2:D10)

LOOKUP is an old deprecated Excel function but it is still proposed in many Excel forums as a solution because of its behaviour of returning the last match when there are multiple matches. Today we would use XLOOKUP but no assistant proposed that because it has as yet a low frequency in the LLM data sources. But if you ask to use XLOOKUP, they do provide formulas using it. My XLOOKUP formula is
=XLOOKUP(H2,(B2:B10=G2)*(C2:C10)*(C2:C10<=H2),D2:D10, -1, -1) using backwards search.

The Boolean multiplications create an array {0;1;0;0;1;0;0;0;0}. Because LOOKUP requires the array to be sorted but ignores error values, the zeroes are changed to #DIV/0! by dividing 1 by the values. LOOKUP then just sees a series of 1s and uses the position of the last one found as an index into the prices.

Repeating ChatGPT-3.5 for interest produced yet another version, using AGGREGATE:

=INDEX(D:D,AGGREGATE(14,6,ROW($B$2:$B$10)/($B$2:$B$10=G2)/($C$2:$C$10<=H2),1))

The 14 is the function number for LARGE, 6 means ignore error values, and the rest creates an array of either DIV/0 errors or row numbers where the conditions are met.

Bing at first suggested the incorrect VLOOKUP and INDEX/MATCH formulas described above. It finally offered the LOOKUP(2 version:
=LOOKUP(2,1/((($B$2:$B$10=G2)*($C$2:$C$10<=H2))),$D$2:$D$10).

Exercise #4: trying confusions with the models

Back in exercise #2, I said that when I asked " What is wrong with this Excel formula: =VLOOKUP(J11,$C$10:$B$110,2,FALSE)" it correctly replied " The range $C$10:$B$110 is in the wrong order. It should be $B$10:$C$110."  I wondered how far a probabilistic language model could go in recognising logical or structural inaccuracies. For example, it cannot solve Wordle problems because it works in tokens, fragments of words, rather than letters. I therefore tried it with less common column letters than B or C, which are very common in internet questions and solutions.

| Question | Answer | Assessment |
|---|---|---|
| I first tried columns JAN and FEB. JAN is after FEB in column order so if you enter JAN:FEB Excel will autocorrect it to FEB:JAN. | | |
| What is wrong with this Excel formula: =VLOOKUP(2,Jan:Feb,2,FALSE) | ChatGPT: The range Jan:Feb is not a valid range reference. It should be a specific range such as A1:B10. | Incorrect |
| Same | Bing: There is nothing wrong with the formula itself | Correct |
| Retry with capital letters: What is wrong with this Excel formula: =VLOOKUP(2,JAN:FEB,2,FALSE) | The formula you provided appears to be a valid VLOOKUP formula. However, without more context about the data in the JAN and FEB named ranges…. | |
| I said that JAN & FEB were column letters, not range names. Bing insisted that they were incorrect and even after a long exchange showing that they were columns 6800 and 4188, and it agreed that JAN was between JAM and JAO it still insisted. You can see why people say that Bing's persona is "Fight me". | Bing: In summary, while `JAN` and `FEB` may be valid combinations of letters, they are not valid column references in Excel because they do not follow the base-26 numbering system used by Excel to represent columns. | Incorrect |
| What is wrong with this Excel formula: =VLOOKUP(2,XYC:XYB,2,FALSE) | The range XYC:XYB is in reverse order. | Correct |
| What is wrong with this Excel formula: =VLOOKUP(2,WAY:WAS,2) | The range "WAY:WAS" is not a valid range reference. It should be a specific range of cells, such as "A1:B10". | Incorrect |
| What is wrong with this Excel formula: =VLOOKUP(2,ABC:ABB,2) | The range specified in the formula is incorrect. The range should be specified as ABC:ABB instead of ABB:ABC. | It seems to see that something is wrong but does not know what. |
| What is wrong with this Excel formula: =VLOOKUP(2,FEB:JAN,2,FALSE) | The range FEB:JAN is not in the correct order. It should be JAN:FEB. | It is easily confused by column letters that look like words. |
| | | |

**Lesson 10: These are language models, not logical or mathematical models.**